\documentclass[useAMS,usenatbib]{mn2e}
\usepackage{epsfig,amssymb,amsmath,graphicx,lscape}
\usepackage[figuresright]{rotating}

\title[AstroSat view of `Clocked' burster GS 1826-238]
{AstroSat view of `Clocked' burster GS 1826-238: broad-band spectral nature of persistent and burst emission during soft state}
\author[Vivek K. Agrawal]{Vivek K. Agrawal$^1$$\thanks{E-mail:
vivekag@ursc.gov.in}$, Anuj Nandi$^1$, Tilak Katoch$^2$\\
$^1$Space Astronomy Group, ISITE Campus, U. R. Rao Satellite Center, Bangalore, 560037, India \\
$^2$ Department of Astronomy and Astrophysics, Tata Institute of Fundamental Research, Homi Bhabha Road, Colaba, Mumbai 400005, India
}
\begin{document}


\pagerange{\pageref{firstpage}--\pageref{lastpage}}
\maketitle
\label{firstpage}

\begin{abstract}
In this paper, we have carried out a detailed study of the `Clocked'
burster GS $1826-238$ using $\sim$ 90 ks  broad-band (0.7 - 60.0 keV)
data obtained with {\it AstroSat} observatory. The source was observed
during a soft spectral state and traced a `banana' type track in the
colour-colour diagram (CCD). We find that a combination of   thermal
component (multi-colour disc/bbodyrad) and  Comptonized component
is statistically good description for all the sections of the track in
the CCD.  The corona becomes optically thick ($\tau$ increases from $\sim$
5 to 21) and cooler ($kT_e$ decreases from $\sim$ 4.8 to 2.2 keV) as
the source moves up in the `banana' branch. Probably cooling is caused by
increase in the supply of soft-seed photons from the disc/boundary-layer.
Reflection signature is observed at upper `banana' branch
of the source.  Two type-I X-ray bursts are detected during the {\it
AstroSat} observations. During the  bursts,  hard X-rays increased
unlike previous observations where a reduction in hard X-rays is observed
during the bursts. Decrease in the electron temperature and
increase in the optical depth are observed during the bursts. The PSD (Power Spectral
Density) of all the sections of the CCD  can be represented by a pure
power-law component. The strength of this component increases from $\sim$
1\% to 4.5\% as the source moves up in the `banana' track. Search for
burst oscillations gave a null result.  We discuss the implications of
our results in the context of previous findings.
\end{abstract}


\section{Introduction}
Luminous low mass X-ray binaries (LMXBs) containing an accreting neutron
star (NS) are broadly divided in two groups: Atoll and Z-sources. The
classification is based on the pattern that they trace in colour-colour
diagram (CCD). Z-sources trace a Z-shaped path on the CCD and the Atoll
sources trace a fragmented `C' type pattern on the CCD \citep{Hasvan89}. The
fragmented pattern has two distinct parts `island' state and curved
`banana' state. Z-sources are persistent sources with luminosity
varying in the narrow range of $0.5-1.0$ $L_{Edd}$. However, luminosity
of Atoll sources  varies in the range of $0.01 - 0.1$ $L_{Edd}$.

X-ray spectra of the NS-LMXBs are often modeled as sum of soft and hard
spectral components. The soft component is either modelled with a multi-colour
disk (MCD) emission or a single temperature blackbody (BB) emission. The
hard component is described by  Comptonized emission. Therefore, to model
the X-ray spectra two approaches exist.  In first approach, MCD plus
Comptonized emission model is used to fit the X-ray spectra of the NS-LMXBs
\citep{disalvo00a,disalvo02, agr03, tarana08, agr09} and in the second
approach, spectra of these sources are modelled using sum of  BB and
Comptonized emission \citep{piraino00, piraino07, disalvo00b, disalvo01,
barret02,wang19}. However, it is very difficult to differentiate between
MCD and BB models. The Comptonized component of Z-sources has electron
temperature in the range of $2-4$ keV and optical depth in the range of
$5-15$ \citep{disalvo01,disalvo02, agr03, agr09, agr20a,agr20b}. However,
Atoll sources exhibit two types of state: soft state and hard state
\citep{barret00,barret01,Gierlinski02, tarana08, wang19}. The hard state
is usually associated with `island' branch and the soft state with
`banana' branch. In the hard state, Comptonized corona has temperature
in the range of $10-60$  keV and optical depth $<3$ \citep{barret00,
Gierlinski02, tarana07, tarana08}. However, in the soft state 
Atoll sources exhibit  spectral characteristics similar to Z-sources
\citep{barret00, barret02, Gierlinski02, tarana08, tarana07, agr18}.
The spectral evolution studies  have been carried out for many Atoll
sources \citep{barret02, Gierlinski02, tarana07, tarana08, agr18, wang19}
and Z-sources \citep{disalvo00a, disalvo02, agr03, agr09, agr20a, agr20b}
in order to understand the origin of the path traced by them in the CCD,
dynamics of accretion process and emission mechanisms.

GS $1826-238$ is a `clocked X-ray burster', discovered serendipitously
during GINGA observations \citep{Makino88}. Owing to its similarity with
Cyg X-1, it was initially classified as blackhole candidate. Later,
the detection of the three type-I X-ray bursts with {\it BeppoSAX}
established that the compact object in this system is a  weakly magnetized
NS \citep{Ubertini97}.  The source has exhibited regular type-I X-ray
bursts since its discovery. The source has remained in the persistent hard
spectral state until June 2014 \citep{sordo99, Cocchi11, Asai15}. The
{\it MAXI} and {\it Swift-BAT} observations revealed transition to the
soft spectral state on 2014 June 8 \citep{Asai15, Chenvez16}. The source
remained in the soft state for more than 2-months. \cite{Chenvez16}
fitted the joint  {\it Swift} XRT ($0.3-10$ keV) and {\it NuSTAR} $(3 -
40$ keV) spectrum during the soft state with a double Comptonization
model \citep{Chenvez16}. The temperature ($kT_e$) of soft Comptonized
component was found to be $\sim$ 3 keV while $kT_e$ of the hard Comptonized
component was fixed at  20 keV. The source again made a transition to the
soft spectral state in 2015 (MJD 57220 onwards) (see \citealt{Sanchez20}).

NS-LMXBs (mainly Atoll sources) also exhibit type-I X-ray bursts, caused
by unstable thermonuclear burning of accreted material on the surface of
the weakly magnetized neutron stars. During the thermonuclear burning
supply of the seed photons from the surface of the NS is expected to
increase and cool the Comptonized corona around the NS \citep{Maccarone03,
Ji13, Chen13, Sanchez20}.  Cooling of the corona during the type-I X-ray bursts
has been observed in this source.  During
the type-1 X-ray bursts, a soft X-ray excess was observed which was
modeled with blackbody component \citep{Sanchez20}. A drop in the hard
X-ray flux was also reported during the type-I X-ray burst observed in
this source \citep{Ji14, Sanchez20}. \cite{Chenvez16} detected the Eddington
limited type-I X-ray burst and derived the distance of the source to be
5.7$\pm$0.2 kpc, assuming an isotropic emission. The {\it NICER} observations
of this source revealed $7-9$ mHz oscillations with fractional rms 2\%
at 6 keV \citep{strohmayer18}.

In the present work, we have carried out a detailed broad-band spectral
($0.7-25$ keV) study of the source GS $1826-238$ using the {\it AstroSat}
data. The source was in the soft state during our observations and also
showed two type-I X-ray bursts during the {\it LAXPC} observations.
An increase in the hard X-rays was seen during the bursts. Also,
reflection feature was observed during {\it AstroSat} observations.
The remainder of the paper is organized as follows.  The details of
observations and methods  adopted for the data reduction are given in
$\S$2. The methodology used for data analysis, modelling of the energy
spectra and the power density spectra are presented in $\S$3. 
The results of analysis are also presented in $\S3$. Finally, the
implications of our results and conclusions are discussed in $\S$4.
\begin{figure}
	\includegraphics[width=0.55\textwidth]{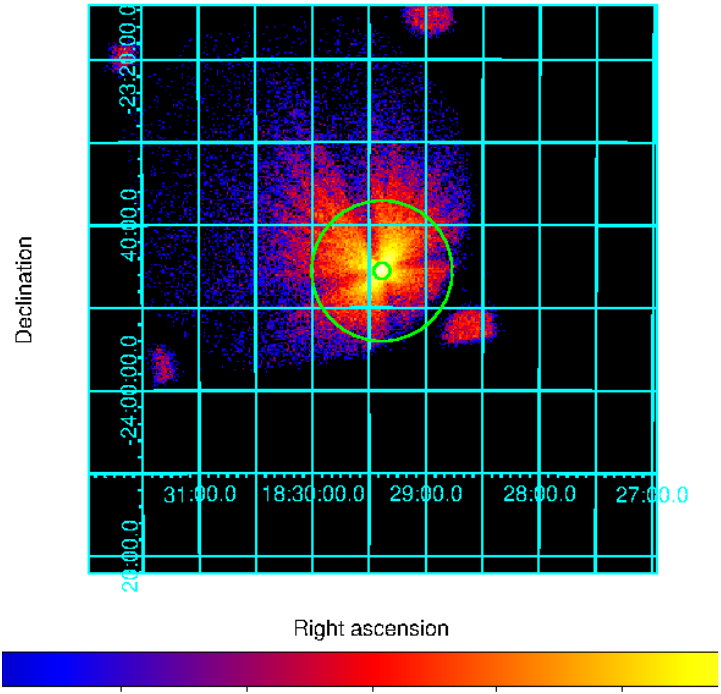}
	\caption{The SXT image in the energy band $0.7-8.0$ keV of the source GS 1826-238. The PC mode data during Obs A has been used to create the image. The four spots at the corners are Fe$^{55}$ sources, used for calibration. The green annulus shows the source selection region. See text for details.}
\end{figure}

\begin{table*}
\caption{Fit statistics for used spectral models to describe the X-ray spectra of Obs A and Obs B.}

\begin{tabular}{lcccc}
\hline
Models & Obs A & Obs B & Remarks\\
 & ($\chi^2$/dof) & ($\chi^2$/dof) & \\
\hline
\hline
{\it nthComp} & 1146/658 & 449/298 & seed DISKBB \\
{\it nthComp+bbodyrad} & 1018/656 & 396/296 & seed DISKBB\\
{\it irefl*nthComp+bbodyrad} & 992/655  & 395/295&seed DISKBB\\
{\it nthComp+diskbb}      & 1008/656 &405/296 &seed BBODY\\
{\it irefl*nthComp+diskbb}   & 997/655  & 401/295& seed BBODY\\
{\it nthComp+nthComp}        &1017/656  & 407/296 & seed DISKBB/BBODYRAD\\
{\it diskbb+bbodyrad}        & 1045/658 & 697/298 & - \\
\hline
\hline
\end{tabular}
\end{table*}

\section{Observations and Data Reduction}

The source GS $1826-238$ was observed twice, from 2016 March 9 to 2016
March 11 (Obs A) and on 2016 August 17 (Obs B) for a total good time
of $\sim$ 90 ks, with {\it AstroSat} instruments: Soft X-ray Telescope
({\it SXT}) and Large Area X-ray Proportional Counter ({\it LAXPC}).
{\it SXT} consists of a focusing optics and a CCD imager. It provides
spectroscopic and temporal measurements in the energy range $0.3-8.0$ keV
\citep{singh16}. {\it LAXPC} is a gas proportional counter and operates
in the $3-80$ keV energy band \citep{Yadav16}. There are three {\it
LAXPC} units: LAXPC10, LAXPC20 and LAXPC30 with combined effective
area of 6000 cm$^2$. {\it SXT} operates in the photon counting (PC)
and the fast window (FW) modes. In the PC mode, data from the entire
CCD is read out. Time resolution in this mode is 2.38 s. In the FW mode,
the data from the central 150 $\times$ 150 pixels are collected. The
read out time in this mode is 278 ms. {\it LAXPC} operates in the event
analysis and the broad-band counting mode. In the event analysis mode,
the arrival time  of each photon is tagged with an accuracy of 10
$\mu$s. In the broad-band counting mode, one can select the binsize
from 16 ms to 2048 ms. During our observations, the  data from {\it
SXT} was collected using the PC mode (Obs A) and FW mode (Obs B), and
the event analysis mode was used to collect data from the {\it LAXPC}
(for both Obs A and Obs B).  \begin{figure}
\includegraphics[width=0.55\textwidth,angle=-90]{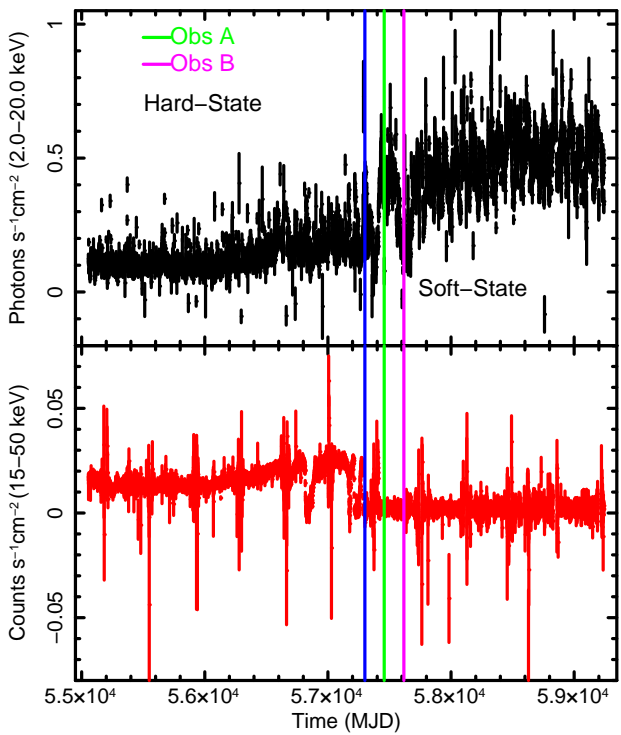}
	\caption{The upper panel shows the {\it MAXI} lightcurve in the energy range $2-20$ keV and the bottom panel  shows {\it Swift-BAT} count rate variations in the energy range $15-50$ keV. The vertical blue line divides the lightcurve in two regions namely hard-state (left) and soft-state (right). We also mark the {\it AstroSat} observation epochs by  green (Obs A) and magenta (Obs B) lines. }
\end{figure}

\begin{figure}
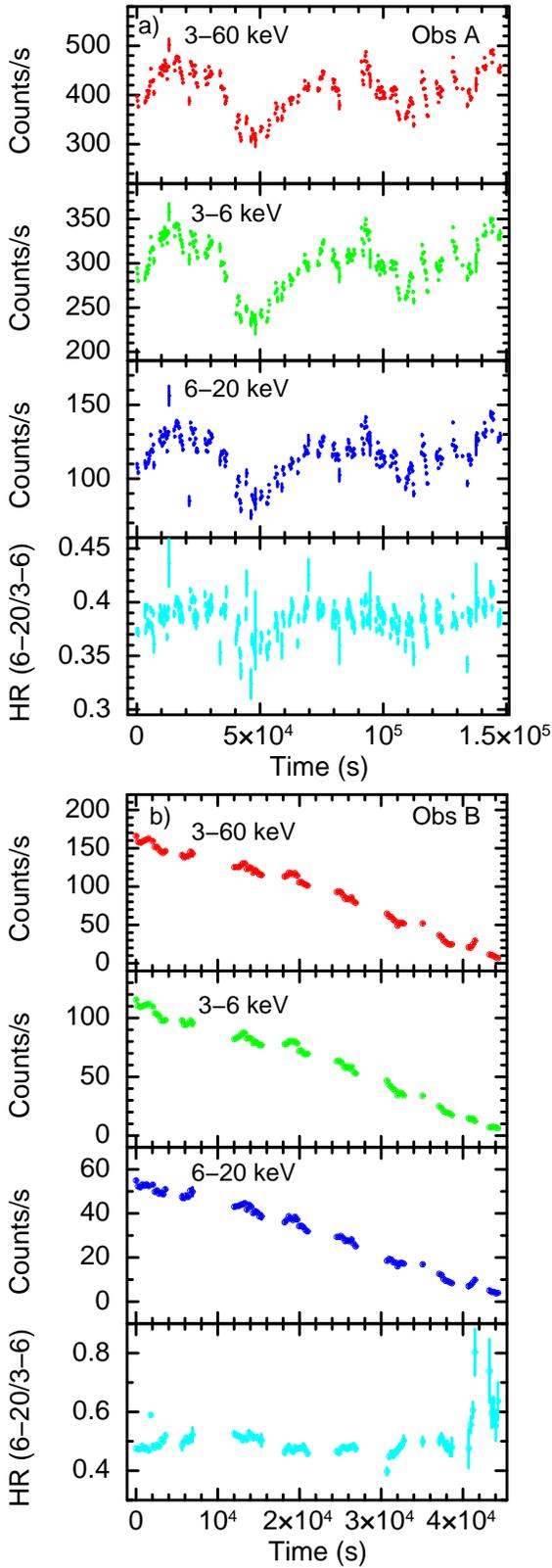

\includegraphics[width=0.6\textwidth,angle=-90]{Lightcurve-obs1.eps}
\includegraphics[width=0.6\textwidth,angle=-90]{Lightcurve-obs2.eps}
\caption{ Lightcurves of the source GS $1826-238$: (a) for Obs A and (b) for Obs B. The lightcurve is created in the energy bands $3-60$ keV, $3-6$ keV and $6-20$ keV using binsize of 256 s. The hardness ratio is shown in the bottom panels of the lightcurves and defined as the ratio of the count rates in the energy band $6-20$ and $3-6$ keV.} 
\end{figure}
\begin{figure}
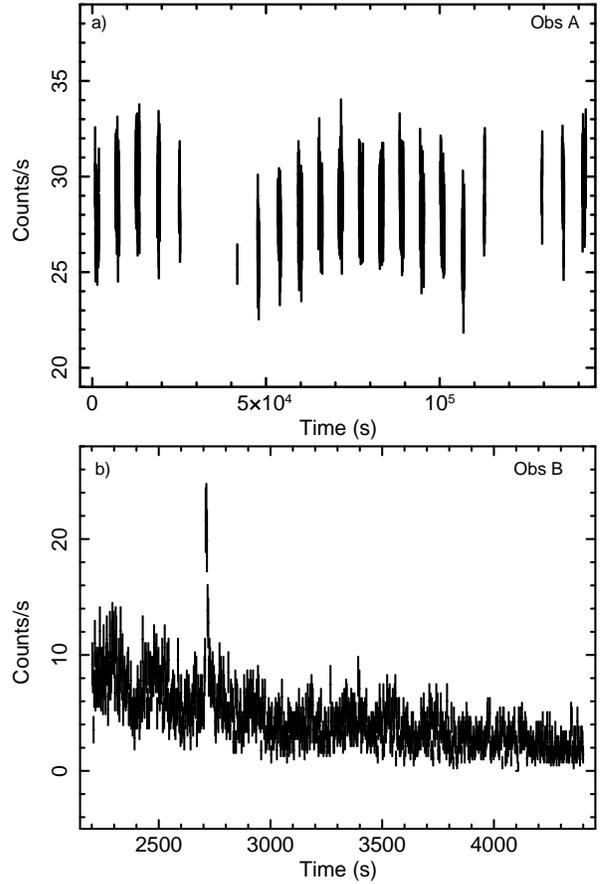

\includegraphics[width=0.33\textwidth,angle=-90]{gs1826-src-sxt-366.eps}
\includegraphics[width=0.33\textwidth,angle=-90]{gs1826-src-sxt-4802.eps}
\caption{SXT lightcurves of the source GS $1826-238$.  The lightcurve is created in the  $0.7-8$ keV energy band.  Obs A lightcurve is created using PC mode data and bin size of 23.8 s. Obs B lightcurve is created using FW mode data and bin size of 2.8 s. A type-I X-ray burst was detected during Obs B.} 
\end{figure}
\begin{figure}
\includegraphics[width=0.33\textwidth,angle=-90]{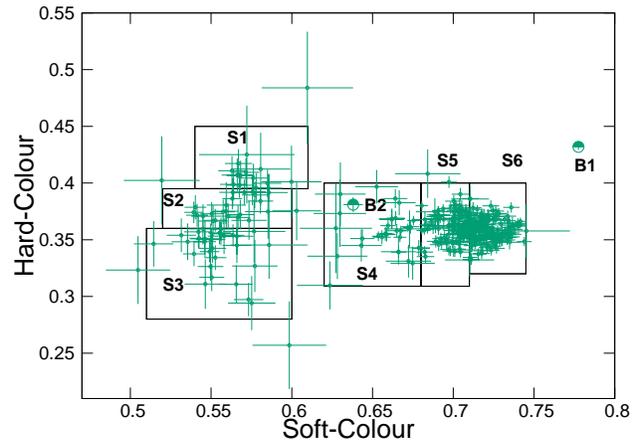}
\caption{ Colour-colour diagram of GS 1826-238 (see text for details). The different sections of the CCD have been marked with label S1-S6. The points corresponding to the bursts are marked with B1 and B2 and displayed using open circle symbol.}

\end{figure}
 We use the latest version of  {\it LAXPC} data analysis software
 \textbf{LaxpcSoft}\footnote{http://www.tifr.res.in/$\sim$astrosat\_laxpc/LaxpcSoft.html},
 provided by the LAXPC team  and follow the data reduction procedure
 given in the readme file (see also \citealt{agr18}; \citealt{agr20b}). We
 reduce level-1 data products to get final LAXPC lightcurves and spectra.
 \textbf{XSELECT} version 2.4d is used to generate the image, spectra
 and the lightcurves from the {\it SXT} level-2 data. The {\it SXT}
 image in the energy range $0.7-8.0$ keV (for Obs A) is shown in the
 Figure 1.  The $0.5-8.0$ keV count rate of the source during Obs
 A exceeded the pileup limit ($>$ 40 cps). Hence, we use an annular
 region with  inner radius of $1\arcmin$ arcmin and an outer radius
 of $12\arcmin$. We show the annular extraction region in
 Figure 1 with two green circles corresponding to inner and outer
 radius of region of selection. However for FW mode (Obs B),  we use a
 circular $8\arcmin$ region to extract the spectra and lightcurves. The
 background spectra obtained from the dark sky observations provided
 by the instrument team were used to subtract the background. The
 response matrices for both {\it LAXPC} and {\it SXT} were provided by
 the instrument team. We create ARF files for {\it SXT} using task {\it
 sxtARFModule}\footnote{http://www.tifr.res.in/$\sim$astrosat\_sxt/dataanalysis.html}.

\section{Data Analysis and Modelling}

\subsection{Lightcurves and CCD}
In Figure 2, we show the {\it MAXI} (top panel) lightcurve in the energy
band $2-20$ keV and {\it Swift-BAT} lightcurve in the energy band $15-50$
keV (bottom panel). We also mark the time of {\it AstroSat} observations
by green (Obs A) and magenta (Obs B) line. The MAXI count rate increased
after MJD 57220 (2015 July 17). At the same time, a drop in the $15-50$
keV flux was observed. This suggested that source made hard to soft state
transition near MJD 57220. We mark the time of the state transition by
blue line.  Figure 2 suggests that the {\it AstroSat} data are obtained
in the soft state of the source. However, during Obs B, $2-20$ keV flux
is less compared to Obs A (see Figure 2). We use the {\it LAXPC10}
data to create the lightcurves in the energy band $3-60$ keV, $3-6$ keV
and $6-20$ keV for Obs A and Obs B. The bin size used is 256 s. We  show
these lightcurves along with the  hardness ratio (the ratio of count rates
in the energy bands $6-20$ keV and $3-6$ keV) in Figure 3a and Figure 3b.
 Two type-I X-ray bursts (B1 during Obs A and B2 during Obs B) were recorded during our observations.

We also create the lightcurves
using the {\it SXT} data in the energy band $0.7-8.0$ keV. One type-I
X-ray burst was recorded during Obs B with {\it SXT}. The {\it SXT}
lightcurves are shown in the Figure 4a and 4b.  We create {\it LAXPC}
lightcurves with 256 s binsize in 4 energy bands: $3.0-4.5$ keV,
$4.5-6.5$ keV, $6.5-10.0$ keV and $10-18$ keV.  These lightcurves
are used to create the colour-colour diagram (CCD, see Figure 5). Here,
the soft colour is defined as the ratio of count rates in the energy
bands $4.5-6.5$ keV and $3.0-4.5$ keV, and hard colour is the ratio of
count rates in the $10-18$ keV and $6.5-10.0$ keV band. The pattern
traced by the source in the CCD  is reminiscent of the English letter
`C'. We divide the pattern into 6 parts (S1, S2, S3, S4, S5 and S6) to
investigate the evolution of broad-band spectral and temporal properties.
Figure 6 shows the  X-ray bursts (B1 and B2) detected using {\it LAXPC}
in the energy bands: $3-60$ keV (top panel), $3-20$ keV (second panel)
and $20-60$ keV (third panel) for Obs A and Obs B. The hardness ratio, 
defined as the ratio of the count rates in the energy bands $20-60$ keV and $3-20$ keV is shown in the bottom panels.

The average $3-60$ keV count rate of the persistent emission during Obs A
was $\sim$ 400 counts/s and during Obs B was $\sim$ 80 counts/s. During
Obs B,  the count rate systematically decreased from $\sim$ 160 counts/s to
10 counts/s (see Figure 3). The peak $3-60$ keV  count rate of the X-ray burst observed
during Obs A was 6200 counts/s and that during Obs B was around 3960
counts/s. We note that bursts are detected in the $20-60$ keV band as well
(see Figure 6) and the peak count rate was 86 counts/s for the burst
detected during Obs A and it was 55 counts/s for the burst detected
during Obs B. Only burst B2 was detected during {\it SXT} observations
(see Figure 4b). The soft colour varies between  0.5 to 0.8 and the
hard colour varies between 0.25 to 0.45 (see Figure 5). The burst B2
was detected at lower soft colour values compared to the burst B1.
\begin{figure}
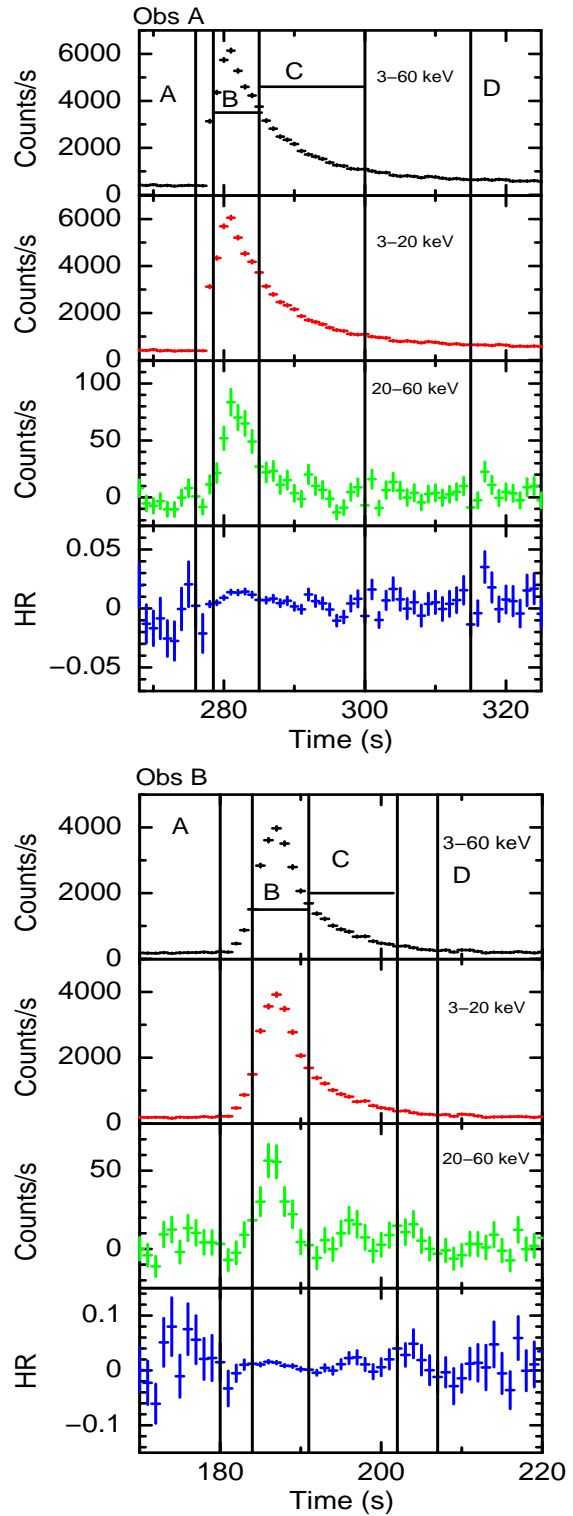

\includegraphics[width=0.57\textwidth, height=0.33\textheight, angle=-90]{burst1-plot-new-sel-20keV.eps}
\includegraphics[width=0.57\textwidth,height=0.35\textheight, angle=-90]{burst2-plot-new-sel-20keV.eps}
\caption{The type-I X-ray bursts observed during the  Obs A ( B1) and Obs B (B2). The lightcurves in the three energy bands $3-60$ keV, $3-20$ keV and $20-60$ keV are shown for both bursts. We also provide the hardness ratio (HR) which is the ratio of count rates in the $20-60$ and $3-20$ keV bands. Segments A, B, C and D correspond to the pre-burst, peak, decay and post-burst time respectively. Since we have expanded the burst lightcurves, full pre-burst and post-burst intervals are not shown here and exact time intervals used are given in the Table 5.}
\end{figure}
\begin{figure*}
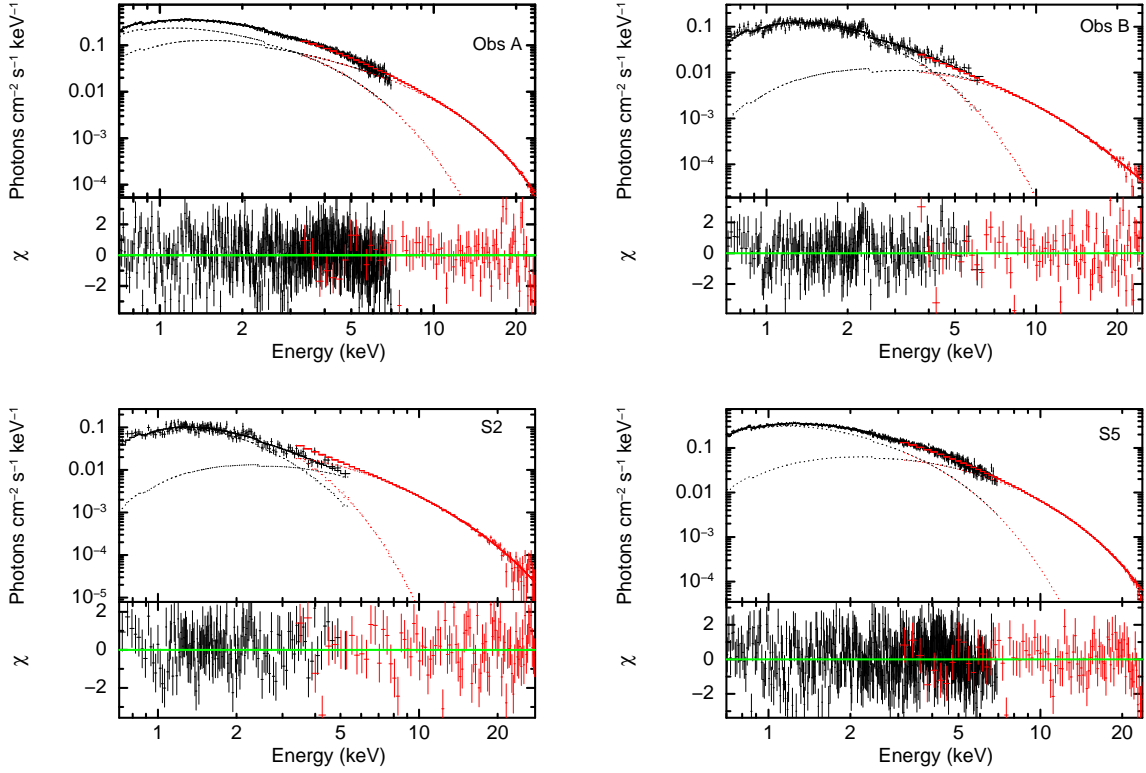

\centering
\begin{tabular}{@{}cc@{}}
\includegraphics[width=0.29\textwidth,angle=-90]{mcd-nthcomp-edge-total-irefl.eps}&
\includegraphics[width=0.29\textwidth, angle=-90]{diskbb-seedncd-nthcomp-edge-irefl.eps} \\
\includegraphics[width=0.29\textwidth,angle=-90]{s2-nthcomp-diskbb-final-edge-seedbb-irefl.eps} & 
\includegraphics[width=0.29\textwidth,angle=-90]{s5-nthcomp-diskbb-final-seedbb-edge-irefl-new1.eps} \\
\hspace{-1in}
\end{tabular}
	\caption{ The unfolded X-ray spectra for Obs A and Obs B along with the section S2 and S5 are shown. The fitted model is {\it TBabs*edge*(irefl*nthComp+diskbb)}. The bottom panels of the figures show residuals in units of sigma.}
\end{figure*}
\begin{figure}
\includegraphics[width=0.6\textwidth,angle=-90]{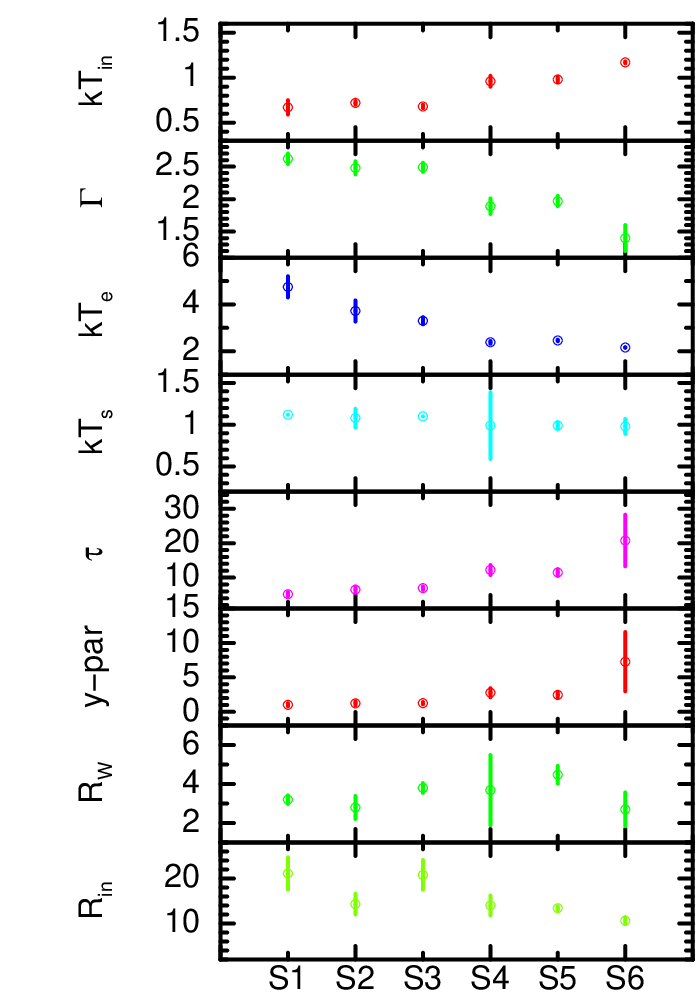}
	\caption{ The variation of best-fit X-ray spectral parameters along the CCD for model {\it TBabs*edge*(irefl*nthComp+diskbb)}. See text for details}
\end{figure}
\begin{figure}
\includegraphics[width=0.6\textwidth,angle=-90]{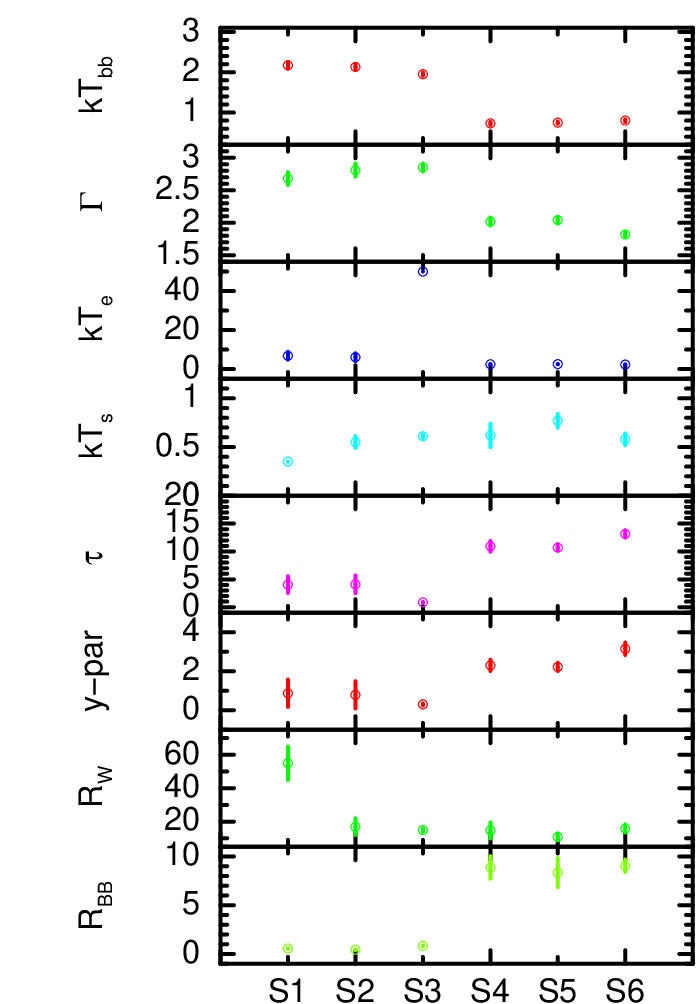}
	\caption{ The variation of best-fit X-ray spectral parameters along the CCD for model {\it TBabs*edge*(irefl*nthcomp+bbodyrad)}. See text for details}
\end{figure}
\begin{table*}
	\caption{ The best fit results, obtained using  model ({\it TBabs*edge*(irefl*nthComp+diskbb)}).  $N_H$ is equivalent hydrogen column density in units of 10$^{21}$ cm$^{-2}$. $kT_{in}$ and $N_D$ are inner disc temperature and normalization of {\it diskbb} respectively. $\Gamma$, $kT_e$ and $N_{Th}$ are photon index, electron temperature and normalization of Comptonized component respectively.  $f_d$ and $f_{th}$ are disc and Comptonization fluxes respectively in the energy range $0.3-50.0$ keV. Fluxes are in units of 10$^{-9}$ $ergs/s/cm^2$. $\tau$, $y-par$ is optical depth and Comptonization y-parameter. $R_{in}$ is the inner disc radius and $R_W$ is the  Wien radius. $refl\_frac$ is reflection fraction and $i$ is inclination angle.}
	\begin{tabular}{ccccccc}
\hline
Parameters & S1 & S2 & S3 & S4 & S5 & S6\\
\hline
$N_H$      &4.0(fix) & 4.0$\pm$0.1 &4.0$\pm$0.4 &2.7(fix) &2.7(fix) & 2.7(fix) \\
$kT_{in}$ (keV) & 0.67$\pm$0.08  & 0.72$\pm$0.02 &0.68$\pm$0.02 &0.96$\pm$0.06  & 0.96$\pm$0.04&1.17$\pm$0.02\\
$N_{D}$         &360.5$\pm$55.5&164.25$\pm$25.5 &345$\pm$51&157.5$\pm$26.5&143.5$\pm$14.5&90.5$\pm$2.6\\
$\Gamma$        &2.62$\pm$0.08 &2.48$\pm$0.10&2.49$\pm$0.07&1.89$\pm$0.12&1.97$\pm$0.08 & 1.40$\pm$0.3 \\
$kT_e$ (keV)          &4.75$\pm$0.45 & 3.72$\pm$0.45 &3.32$\pm$0.15&2.38$\pm$0.06&2.44$\pm$0.04&2.16$\pm$0.03\\
$kT_s$ (keV)         &1.12(fix)     & 1.08$\pm$0.11 &1.1(fix) & 0.99$^{+0.48}_{-0.23}$ & 0.95$\pm$0.12&0.98$\pm$0.10\\
$N_{Th}$        & 0.024$\pm$0.005  &0.015$\pm$0.002 &0.025$\pm$0.01&0.045$\pm$0.01 &0.060$\pm$0.01 & 0.03$\pm$0.01\\
$refl_frac$     &$<$0.05 & $<$0.03 &$<$0.015  & 0.080$\pm$0.03 & 0.075$\pm$0.01 & 0.12$\pm$0.02\\
$\cos i$        & 0.3(fix) & 0.3 (fix) & 0.3(fix) &0.3(fix)&0.3(fix) &0.3(fix) \\

\hline
\hline
 & & & Derived Values & & &\\
\hline
$f_d$  & 1.42   &0.82$\pm$0.04 & 1.61$\pm$0.09 & 2.68$\pm$0.43 & 2.80$\pm$0.21&3.68$\pm$0.43\\       
$f_{th}$ & 0.97& 0.51$\pm$0.05 & 0.81$\pm$0.06 & 1.67$\pm$0.43& 2.25$\pm$0.21 & 1.92$\pm$0.22\\
$\tau$         &5.18$\pm$0.65 & 6.52$\pm$0.8 & 6.95$\pm$0.51 & 12.20$\pm$1.48 & 11.49$\pm$0.97 & 20.75$\pm$7.5\\
$y-par$        &1.02$\pm$0.23 &1.24$\pm$0.34  &1.25$\pm$0.18 & 2.77$\pm$0.66 & 2.45$\pm$0.42 & 7.28$\pm$4.35 \\
$R_{in}$ (km)      &21.13$\pm$3.55 &14.35$\pm$2.33 & 20.82$\pm$3.35 & 14.03$\pm$2.20 &13.45$\pm$0.45 & 10.65$\pm$0.77\\
$R_W$    (km)      & 3.20$\pm$0.22 & 2.80$\pm$0.59 &3.8$\pm$0.24 &3.69$\pm$1.80 & 4.48$\pm$0.45 & 2.70$\pm$0.87 \\
\hline
\hline
$\chi^2/dof$ &65/87 &256/223 & 273/180 & 427/377 &750/606 & 797/626\\
\hline
\hline
\end{tabular}
\end{table*}
\begin{table*}
	\caption{ The spectral parameters obtained using  model {\it TBabs*edge*(irefl*nthComp+bbodyrad)}. $kT_{bb}$ and $N_{BB}$ are the temperature and normalization of the blackbody component respectively. $f_{bb}$ is the blackbody flux in units of $10^{-9} ergs/s/cm^2$. The $R_{BB}$ is the blackbody radius.}
\begin{tabular}{ccccccc}
	\hline
Parameters & S1 & S2 & S3 & S4 & S5 & S6\\
	\hline
	\hline
$N_H$ & 3.0(fix) &4.0$\pm$0.02 & 4.5$\pm$0.03 & 0.26$\pm$0.03 & 0.25$\pm$0.02 & 0.26$\pm$0.03 \\
$kT_{bb} (keV)$& 2.17$\pm$0.08 & 2.13$\pm$0.06 & 1.95$\pm$0.04 & 0.73$\pm$0.04 & 0.75$\pm$0.03 & 0.80$\pm$0.01\\
$N_{BB}$ & 0.85$\pm$0.14  & 0.59$\pm$0.09 & 2.15$\pm$0.06 & 233$\pm$60 & 214$\pm$80 & 251$\pm$30 \\
$\Gamma$ & 2.68$\pm$0.12 & 2.81$\pm$0.13 & 2.85$\pm$0.06 & 2.02$\pm$0.06 & 2.04$\pm$0.05 & 1.82$\pm$0.04\\
$kT_e$ (keV)  & 6.82$^{+4.01}_{-1.55}$ & 6.05$^{+4.71}_{-1.33}$ & 50.0(fix) & 2.45$\pm$0.06 & 2.48$\pm$0.03 & 2.30$\pm$0.02\\
$kT_s$  (keV) &0.35(fix) & 0.55$\pm$0.06 & 0.61$\pm$0.02 & 0.62$\pm$0.12 & 0.77$\pm$0.07 & 0.58$\pm$0.06\\
$N_{nth}$ & 0.64$\pm$0.05 & 0.26$\pm$0.02 & 0.47$\pm$0.02 & 0.49$\pm$0.07 & 0.50$\pm$0.04 & 0.49$\pm$0.02\\
$refl_frac$ & $<$0.11 & $<$0.08 & $<$0.03  &0.085$\pm$0.03 & 0.095$\pm$0.02 & 0.135$\pm$0.02 \\
$\cos i$    & 0.3(fix) & 0.3(fix) & 0.3(fix) & 0.3(fix) & 0.3(fix) & 0.3(fix)\\
\hline
& & & Derived Values & & &\\
\hline
$f_{bb}$ & 0.16 & 0.12$\pm$0.01 & 0.34$\pm$0.04 & 0.96$\pm$0.09 & 0.89$\pm$0.08 & 1.28$\pm$0.11\\
$f_{nth}$ & 2.65 & 1.65$\pm$0.05 & 1.38$\pm$0.04 & 3.60$\pm$0.25 & 4.58$\pm$0.21 & 4.05$\pm$0.22 \\
$\tau$ &4.05$^{+2.24}_{-0.81}$ & 4.09$^{+2.54}_{-0.91}$ & 0.88$\pm$0.07 & 10.92$\pm$0.95 & 10.70$\pm$0.55 & 13.15$\pm$0.65\\
$y-par$ &0.87$^{+1.12}_{-0.35}$ & 0.79$^{+1.30}_{-0.34}$ &  0.30$\pm$0.05 & 2.30$\pm$0.39 & 2.22$\pm$0.21 & 3.15$\pm$0.32\\
$R_W$ (km)  &55.0$^{+16.5}_{-5.25}$ &17.2$^{+6.2}_{-4.1}$ & 14.97$\pm$1.17 & 14.75$\pm$4.82 & 10.88$\pm$2.07 & 15.88$\pm$2.42\\
$R_{BB}$ (km) &0.52$\pm$0.03 & 0.43$\pm$0.03 & 0.83$\pm$0.04 & 8.88$\pm$1.16 & 8.31$\pm$1.52 & 9.05$\pm$0.65\\
\hline
\hline
$\chi^2/dof$ & 65/87 & 259/223 & 274/181 & 425/376 &748/605 & 792/625\\
\hline
\hline
\end{tabular}
\end{table*}
\begin{table*}
	\caption{ The spectral parameters obtained using  model {\it TBabs*(nthComp+nthComp)}. $\Gamma_1$ and $\Gamma_2$ are photon indices for two Comptonized  components used here. $kT_{e1}$ and $kT_{e2}$ are electron temperatures. $kT_{s1}$ and $kT_{s2}$ are seed photon temperatures. $N_{nth1}$ and $N_{nth2}$ are normalizations. $\tau_1$ and $\tau_2$ are the optical depths for first and second Comptonized component respectively.}
\begin{tabular}{ccccccc}
	\hline
Parameters & S1 & S2 & S3 & S4 & S5 & S6\\
	\hline
	\hline
	$\Gamma_1$ &2.43$\pm$0.21& 2.53$\pm$0.11 & 3.33$\pm$0.3 & 1.89$\pm$0.15 & 1.65$\pm$0.20 & 1.75$\pm$0.15\\
	$kT_{e1}$ (keV) & 4.15$\pm$0.25& 3.80$\pm$0.20 & 5.82$^{+3.22}_{-1.25}$ & 2.42$\pm$0.05 & 2.38$\pm$0.03 &  2.30$\pm$0.01\\
	$kT_{s1}$ (keV)  & 1.26$\pm$0.27& 1.23$\pm$0.08 & 1.20$\pm$0.08 &1.03$\pm$0.29 & 1.03$\pm$0.21 & 0.76$\pm$0.05 \\
	$\Gamma_2$ & 6.33 (fix) & 6.22(fix)& 9.2(f) & 7.55(fix) & 6.1(fix) & 8.0(fix)\\
	$kT_{e2}$ (keV) & 20(f)& 20(f) & 20(f) & 20(f) & 20 (f)& 20(f) \\
	$kT_{s2}$ (keV) & 0.30(f)& 0.66$\pm$0.02 & 0.52$\pm$0.02 & 0.92$\pm$0.12  & 1.02$\pm$0.08& 0.88$\pm$0.04\\
	$N_{nth1}$ ($\times 10^{-2}$) &11$\pm$1.1 & 0.87$\pm$0.1 & 1.72$\pm$0.3 &3.70$\pm$0.8 & 3.14$\pm$0.9 & 9.85$\pm$1.0\\
	$N_{nth2}$ &0.96$\pm$0.05 & 0.25$\pm$0.02 & 0.28$\pm$0.02 & 0.59$\pm$0.01 & 0.62$\pm$0.01 & 0.52$\pm$0.01\\
\hline
\hline
 &  & & Derived Values & & & \\
\hline

	$\tau_1$ & 6.27$\pm$1.12 & 6.26$\pm$0.65 & 3.35$\pm$1.33 & 12.09$\pm$2.12 & 15.02$\pm$4.75 & 14.15$\pm$0.20\\
	$\tau_2$ & 0.49 & 0.50 & 0.27 & 0.37 &  0.52 & 0.70\\
        $y-par1$ &  1.27$\pm$0.23 & 1.16$\pm$0.15  &  0.50$\pm$0.35 & 2.77$\pm$0.97 & 4.20$\pm$2.56 & 3.58$\pm$0.12\\
        $y-par2$ & 0.04& 0.04 & 0.01 & 0.02 & 0.04 & 0.08\\
$R_{W1}$ (km) & 2.58$\pm$0.83 & 1.68$\pm$0.21 & 2.58$\pm$0.4 & 3.11$\pm$1.21 & 2.48$\pm$1.12 & 6.82$\pm$0.95 \\
$R_{W2}$ (km) & 71 & 11.95$\pm$1.18 & 22.89$\pm$2.05 &10.92$\pm$2.95 & 9.86$\pm$1.59 & 11.77$\pm$1.16\\
        $f_{th1}$ &1.3& 0.47$\pm$0.11&0.72$\pm$0.04 & 1.46$\pm$0.33 &1.23$\pm$0.23 & 2.43$\pm$0.14 \\
        $f_{th2}$ & 1.5 & 0.96$\pm$0.1 & 1.33$\pm$0.07  & 2.98$\pm$0.35 & 3.75$\pm$0.14 & 3.06$\pm$0.12\\
	\hline
	\hline
$\chi^2/dof$ &62/88 &  258/224 & 274/181 & 427/377 & 762/606 & 810/626 \\ 
\hline
\hline
\end{tabular}

\end{table*}

\subsection{Modelling of X-ray Spectra}
\subsubsection{Spectral Analysis}
We  create the source and  background spectra for Obs A, Obs B
and different sections of the CCD using the top layer data of LAXPC10. 
We use an annular region  to extract source spectra from SXT PC mode data. To extract source spectra from SXT FW mode data, we use a circular region.
We use the joint {\it SXT} and {\it
LAXPC} spectra for the spectral analysis.
For the spectral fitting, we consider 
$0.7-7.0$ keV energy range for the {\it SXT} and $3-25$ keV range for the
{\it LAXPC}. Since, there is no data with sufficient statistics beyond 25 keV, we restrict our analysis to the energy $<25$ keV.
We use XSPEC version 12.11.1 to carry out the spectral fitting and
 add 1\% systematic error while fitting to account for uncertainty
 in the spectral response.

We attempt various single and multi-components models to fit the X-ray spectra
corresponding to Obs A and Obs B of the source.  We begin with an absorbed
Comptonization model.  To account for Galactic absorption, we use the
XSPEC model {\it TBabs} \citep{wilms00}. We use the Comptonization model
{\it nthComp} \citep{Zdz96} of XSPEC which provides choice to select
the disc blackbody or blackbody as the seed photon spectrum. We select 
the shape of the seed photon spectrum as {\it diskbb}. The reduced $\chi^2$ ($\chi^2_{red}$)
is given in the Table 1. We find that absorbed Comptonization model
provides poor description of the data. Then, we add a ${\it diskbb}$ or
a ${\it bbodyrad}$ component to the spectra.  It is also worth mentioning that there is a gold edge at 2.41 keV in the SXT spectra.  Hence, we add an absorption edge ({\it edge} in XSPEC) at 2.41 keV.
  While fitting the data with
${\it TBabs*edge*(nthComp+diskbb)}$  model, we choose the shape of seed photon
spectrum as ${\it blackbody}$ and the shape of the seed photon spectrum is
taken as ${\it diskbb}$ for the model  ${\it TBabs*edge*(nthComp+bbodyrad)}$.
The fit is significantly improved in both cases, however residual
is found in  the spectral fit of Obs A. We add an ionized reflection
component {\it ireflect} \citep{magdziarz95} to both models.   Since it was difficult to
constrain all the parameters while using the reflection component, we
fix the value of the ionization parameter, the disc inclination and disk
temperature at the best fit values ($\xi = 800$, $\cos i = 0.3$ and $T_{disk} = 0.5$ keV). ${\it TBabs*edge*(irefl*nthComp+diskbb)}$  and ${\it
TBabs*edge*(irefl*nthComp+bbodyrad)}$ models are applied to both spectra
(Obs A, Obs B). We find that addition of the reflection component
improves the fitting of the spectrum for Obs A (see Table 1). However,
for Obs B, an addition of reflection component does not improve the
fit significantly and we could set only a upper limit on reflection fraction ($refl\_frac < 0.04$). The unfolded spectra of Obs A and Obs B with the best fit models ({\it TBabs*edge*(irefl*nthComp+diskbb}) are shown in Figure 7. A double Comptonization model $({\it TBabs*edge*(nthComp+nthComp)})$ also provides an acceptable fit to the spectra.

 We applied these three models to the spectra corresponding to the
 different sections of the CCD. Note that a  combination of {\it diskbb}
 and {\it bbodyrad} models gives a poor fit to the spectra (see Table
 1). The sections S1, S2 and S3 corresponds to Obs B and S4-S6 corresponds
 to Obs A. The reflection component is very weak for sections S1-S3
 and we could find only an upper limit on reflection fraction. However,
 reflection component is significant for the CCD sections S4-S6. Errors
 on the best fit parameters are quoted at 68\% confidence level.

We divide the bursts B1 and B2 into four parts: pre-burst (A), peak (B), decay (C) and
post-burst (D) phases (see Figure 6). The B1 spectra during these different
parts can be completely described by a single component model ({\it
TBabs*(nthComp)}).  An addition of {\it diskbb} component to the spectra
of the pre-burst, peak, decay  and post-burst phases do not improve
the fit. The disc component is not observed because low-energy data
from the {\it SXT} is not available for this burst.  For the burst B2,
a {\it diskbb} component is required during the pre-burst and post-burst
phases. However, {\it diskbb} or {\it bbodyrad} component is not needed
during the peak and decay phases.  Since the seed photons during the type-I
 X-ray bursts will mainly arise in the hot surface of neutron star,
we assume a blackbody shape for the seed photon spectrum.
  However, a  small contribution of seed photons during
the bursts are expected from the accretion disc as well.

Alternatively, we also subtracted the pre-burst spectra from the
peak and  decay spectra for the bursts B1 and B2.  Fitting the resultant
spectra with {\it TBabs*bbodyrad} model provided a poor fit for the
peak ($\chi^2/dof$ = 302/90) and decay part ($\chi^2/dof$ = 248/90) of
the burst B1. Similarly, the above model results into $\chi^2/dof$ = 134/83
and $\chi^2/dof$ = 106/83 for the peak and decay part of the burst B2 respectively.

\subsubsection{Spectral Evolution of Persistent Emission} 

 We list the best fit spectral parameters of model {\it TBabs*edge*(irefl*nthComp+diskbb)} in Table 2. The unfolded X-ray spectra along with the residual for model {\it TBabs*edge*(irefl*nthComp+diskbb)} are shown in Figure 7. We show  total spectra for Obs A and Obs B in the top two panels of Figure 7 for a comparison. The spectra of lower `banana' branch (S2) and upper `banana' branch (S5) are shown in the bottom two panels of Figure 7.

The inner disk radius in the {\it diskbb} model is given by $R_{in}
= \sqrt{\frac{N_D}{\cos i}} D_{10}$.  We use a source distance
of 5.7$\pm$0.2 kpc \citep{Chenvez16} and a source inclination of
75$^\circ$\citep{fujimoto88} to estimate the values of $R_{in}$. We show
the evolution of the X-ray spectral parameters in Figure 8.  We note that
the source shows different spectral behaviour in the lower banana branch
(sections S1-S3) compared to upper banana branch (S4-S6).  For example
temperature $kT_{in}$ of MCD component is lower ($\sim 0.7$ keV) and the
inner disc radius is higher ($15-21$ km) for sections S1-S3 compared to
S4-S6  where $kT_{in} \sim$ 1 keV and $R_{in}$ is $\sim$ $10-14$ km.

We find the evidence of prominent reflection component in the upper
`banana' branch (S4-S6) and in the Obs A (see Figure 5). The reflection signature is absent
in the lower `banana' branch and the Obs B.  We estimated the optical
depth ($\tau$) of the corona by inverting the relationship between the
spectral index ($\alpha = \Gamma_{nth} -1)$, optical depth  and $kT_e$
given in \cite{Zdz96}.  We also computed the Compton y-parameter using
the relation,

\begin{equation}
y=\frac{4kT_e}{m_e c^2} max(\tau, \tau^2).
\end{equation}

We note that $kT_e$  decreases from $\sim$ 4.8 keV to $\sim$ 2.2 keV as the
source moves from S1 to S6 and at the same time $\tau$ increases from
$\sim$ 5 to $\sim$ 21. The Compton y-par also shows a systematic increase as the source moves from S1 to S6.
We estimated the Wien radius  using formula
(see \citealt{Zand99}),
\begin{equation}
R_W = 3 \times 10^4 D \frac{\sqrt{\frac{f_{th}}{1+y}}}{kT_w^2},
\end{equation}
where $kT_w$ is the seed photon temperature and  $f_{th}$ is the flux
of the Comptonized component in the energy range 0.3 - 50 keV. The Wien radius was found to be in the range of $3-4.5$ km. 

In Table 3,  spectral parameters of  model {\it
TBabs*edge*(irefl*nthComp+bbodyrad)} are listed. The evolution of the
spectral parameters is shown in Figure 9. The radius of blackbody
component is higher for sections S4-S6 compared to that obtained
for sections S1-S3. However, blackbody temperature is around $\sim$
2.0 keV for S1-S3 and is lower ($\sim 0.7-0.8$ keV) for S4-S6.
It is also noted that temperature of Comptonized component is high and optical depth is low in S1-S3 compared to their values observed in S4-S6.
We note
that reflection component is weak or absent in lower `banana' branch
(S1-S3). However, it is strong in the sections S4-S6 (upper banana branch),
where luminosity of the source is higher compared to S1-S3.

In Table 4,  we give the best fit spectral parameters of  the model {\it TBabs*(nthComp+nthComp)}.  We note that the electron temperature of the first Comptonized component is $\sim$ $4-6$ keV in the S1-S3 sections and $\sim$ 2.4 keV in the S4-S6 sections. The temperature of second Comptonized component was fixed to 20 keV. The photon index of first Comptonized component increases as the source moves from S1 to S3 and then again decreases as the source moves from S3 to S6. The photon index of second Comptonized component is fixed at the best fit values and is  found to be in the range of $6-9.5$.

\begin{table*}
	\caption{The best fit spectral parameters for the bursts B1 and B2.  $N_H$ is equivalent hydrogen column density in units of 10$^{21}$ cm$^{-2}$. $kT_{in}$ and $N_D$ are inner disc temperature and normalization of {\it diskbb} respectively. $\Gamma$, $kT_e$, $N_{Th}$ and $kT_W$  are photon index, electron temperature normalization and seed photon temperature  of the Comptonized component respectively. $f_d$ and $f_{th}$ are the disc and Comptonization fluxes and in units of $10^{-9}$ $ergs/s/cm^2$.}  
\begin{tabular}{ccccccccc}
\hline
Parameters & \multicolumn{4}{c}{B1} & \multicolumn{4}{c}{B2}\\
\hline
	   & Pre-burst  & Peak  & Decay  & Post-burst  & Pre-burst & Peak  & Decay & Post-burst \\
	   &(76-276s) & (278-285s) & (285-300s) & (315-512s) & (0-180s) & (184-191s) & (191-202s) & (207-407s)\\
\hline
\hline
	$N_H$       &0.25(f) & 0.25(f) & 0.25(f) & 0.25(f) &0.3$\pm$0.07 & 0.25(f) & 0.22(f) & 0.44$\pm$0.06\\  
	$kT_{in}$ (keV)   & -& -& -& -& 0.77$\pm$0.09 & - & - & 0.83$\pm$0.05\\
     $N_D$       &-&-&-&-&74.6$\pm$41.5 & -&-& 140.6$\pm$63.5 \\
     $\Gamma$ &2.28$\pm$0.04&1.45$\pm$0.01 &1.75$\pm$0.08 &2.26$\pm$0.02& 2.01$\pm$0.11 & 1.31$\pm$0.02 & 1.80$\pm$0.08 & 2.62$\pm$0.27\\
	$kT_e$ (keV)& 2.75$\pm$0.07 &2.69$\pm$0.02 &2.56$\pm$0.07 & 2.51$\pm$0.06& 2.72$\pm$0.21&2.48$\pm$0.03&1.94$\pm$0.07 & 3.27$\pm$0.91\\
$kT_{W}$ (keV) & 0.59$\pm$0.03 & 0.53(fix) & 0.99$\pm$0.03&0.54$\pm$0.04&0.48$\pm$0.08 & 0.38(fix) & 0.75(fix) & 1.4(fix)\\
$N_{Th}$&0.35$\pm$0.02&1.15$\pm$0.02 & 0.37$\pm$0.03 & 0.54$\pm$0.06&0.15$\pm$0.06 & 0.48$\pm$0.12&0.33$\pm$0.05 &0.011$\pm$0.001 \\
\hline 
\multicolumn{9}{c}{Derived Values} \\
\hline
$f_{th}$ ($2-50$ keV) &3.38$\pm$0.08&33.1$\pm$0.38 & 16.59$\pm$0.19 & 3.98$\pm$0.0.04&0.36$\pm$0.06 & 21.37$\pm$0.51& 7.08$\pm$0.86& 0.61$\pm$0.0.06\\
$f_{th}$ ($0.3-50$ keV) -&-&-& - & -& 1.34$\pm$0.10 & 22.78$\pm$0.58 & 8.10$\pm$0.80 & 0.84$\pm$0.13\\
$f_{d}$($2-50$ keV)&- & - & - & - &0.15$\pm$0.08 & - & - &0.27$\pm$0.11 \\ 
$f_{d}$($0.3-50$ keV)& - & -& - & -& 1.02$\pm$0.18 & - & -&1.38$\pm$0.15\\
$\tau$ & 8.67$\pm$0.38 & 17.69$\pm$0.43 & 13.16$\pm$1.49 & 9.26$\pm$0.30 & 10.37$\pm$1.49 & 22.69$\pm$1.51 & 14.67$\pm$1.61 & 6.25$\pm$1.28 \\
$y$-par & 1.62$\pm$0.14 & 6.59$\pm$0.32 & 3.46$\pm$0.78 & 1.68$\pm$0.11&2.29$\pm$0.65 & 9.95$\pm$1.33 & 3.26$\pm$0.72 & 0.99$\pm$0.23\\
$R_W$ (km) & 17.65$\pm$1.98 & 40.20$\pm$1.66 & 10.65$\pm$1.20 & 25.31$\pm$4.10 & 8.22$\pm$2.92 & 50.16$\pm$3.5 &12.41$\pm$ 1.33 & 1.57$\pm$0.2 \\
$R_{in}$ (km) & - & - & -& -& 9.6$\pm$5.16 & - & -& 13.25$\pm$5.65\\
\hline
	$\chi^2/dof$ & 77/89 & 100/89 & 121/89 & 116/89 & 116/112 & 81/81 & 92/81 & 119/113\\ 
	\hline
\end{tabular}
\end{table*}
\subsubsection{Spectral Evolution of the Burst Emission} 
The spectral parameters obtained by fitting the spectra during the X-ray bursts are given in Table 5.
The pre-burst, post-burst and the burst spectra of B1 can be
completely described by absorbed Comptonization model. This may be due to
limited statistics and unavailability of the data in the low energy band ({\it SXT
} data). For the burst B2,  the {\it SXT} data were available. The pre-burst and
post-burst emission for the burst B2 require an additional {\it diskbb}
component. However, the disc component vanishes during the peak of the burst
and decay part of the burst. The parameters and fluxes for the pre-burst,
post-burst, peak of the burst and decay part of the burst are listed in
Table 5. The unfolded spectra for different parts of the bursts are shown in Figure 10 (burst B1) and Figure 11 (burst B2). It is observed that the photon index ($\Gamma$) is lower during the peak
and decay part of the bursts B1 and B2  compared to that during the pre-burst
and post-burst emission. Also note that $kT_e$ is lower during the peak and
decay part of the bursts as compared to that during the pre-burst emission. The
optical depth ($\tau$)  and degree of Comptonization ($y-par$) are also
higher during the peak and decay part of the bursts. We also note that the decay spectra are softer compared to the peak spectra. A comparison between decay and peak spectra of burst B2 is shown in Figure 12.
\begin{figure*}
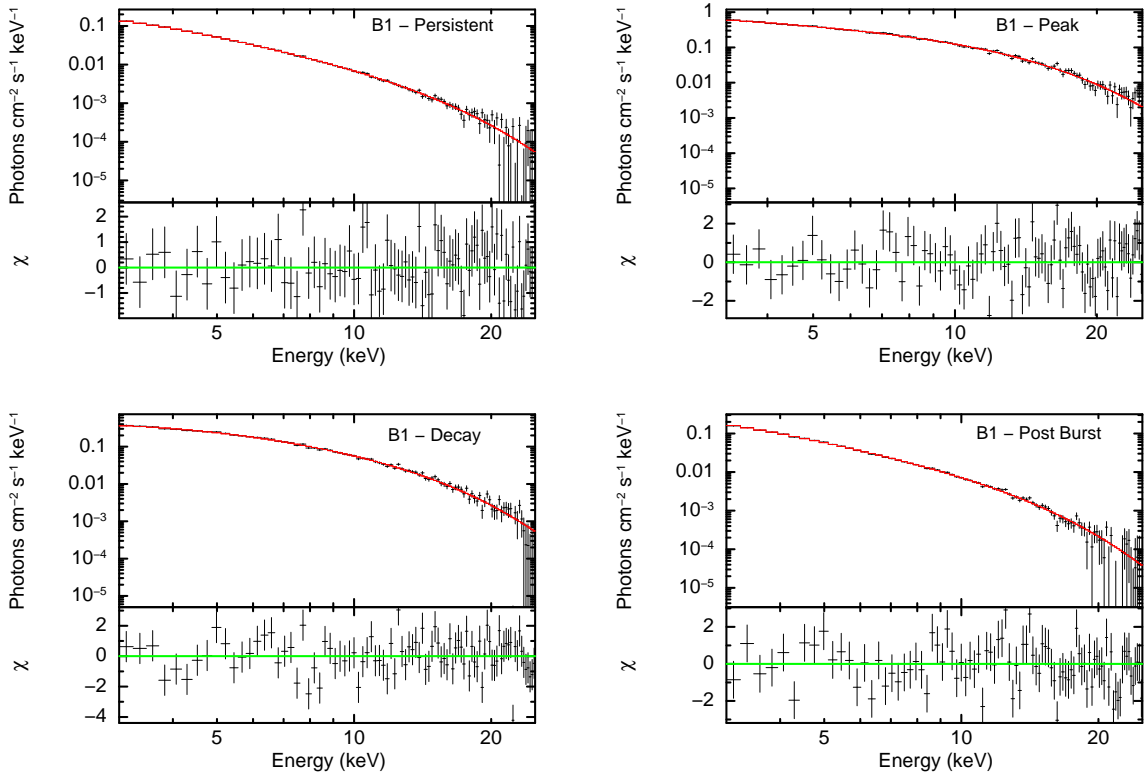

\centering
\begin{tabular}{@{}cc@{}}
\includegraphics[width=0.29\textwidth,angle=-90]{persistent-B1-nthcomp.eps}&
\includegraphics[width=0.29\textwidth, angle=-90]{peak-B1-nthcomp.eps} \\
\includegraphics[width=0.29\textwidth,angle=-90]{decay-B1-nthcomp.eps} & 
\includegraphics[width=0.29\textwidth,angle=-90]{pb-B1-nthcomp.eps} \\
\hspace{-1in}
\end{tabular}
	\caption{The unfolded X-ray spectra for different regions of burst B1. The fitted model is {\it TBabs*(nthComp)}. The bottom panels of the figures show residuals in units of sigma.}
\end{figure*}

\begin{figure*}
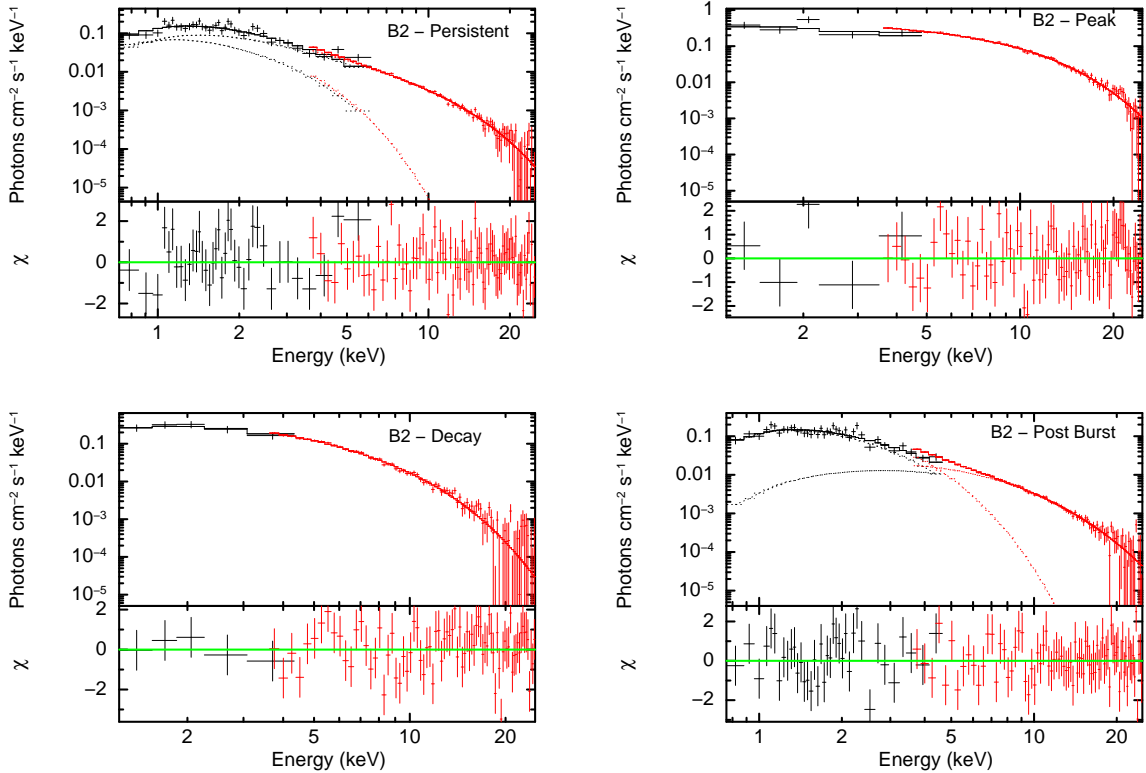

\centering
\begin{tabular}{@{}cc@{}}
\includegraphics[width=0.29\textwidth,angle=-90]{pers-B2-nthcomp-disk.eps}&
\includegraphics[width=0.29\textwidth, angle=-90]{peak-B2-disk-nthcomp.eps} \\
\includegraphics[width=0.29\textwidth,angle=-90]{decay-B2-disk-nthcomp.eps} & 
\includegraphics[width=0.29\textwidth,angle=-90]{pb-B2-disk-nthcomp.eps} \\
\hspace{-1in}
\end{tabular}
	\caption{ The unfolded X-ray spectra for different regions of burst B2. The fitted model is {\it TBabs*edge*(nthComp+diskbb)} for preburst and postburst phase. The model used for peak and decay part of the burst is  {\it TBabs*(nthComp)}. The bottom panels of the figures show residuals in units of sigma.}
\end{figure*}

\begin{figure}
\includegraphics[width=0.29\textwidth,angle=-90]{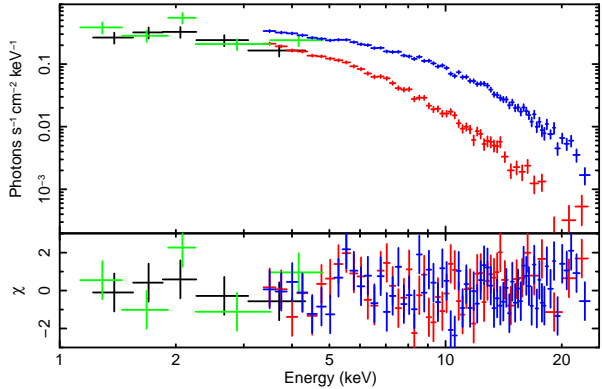}
\caption{ The unfolded X-ray spectra during the peak (green-SXT  and blue-LAXPC) and decay parts (black-SXT and red-LAXPC) of the burst B2. The bottom panel shows residuals in units of sigma.}
\end{figure}
\begin{figure*}
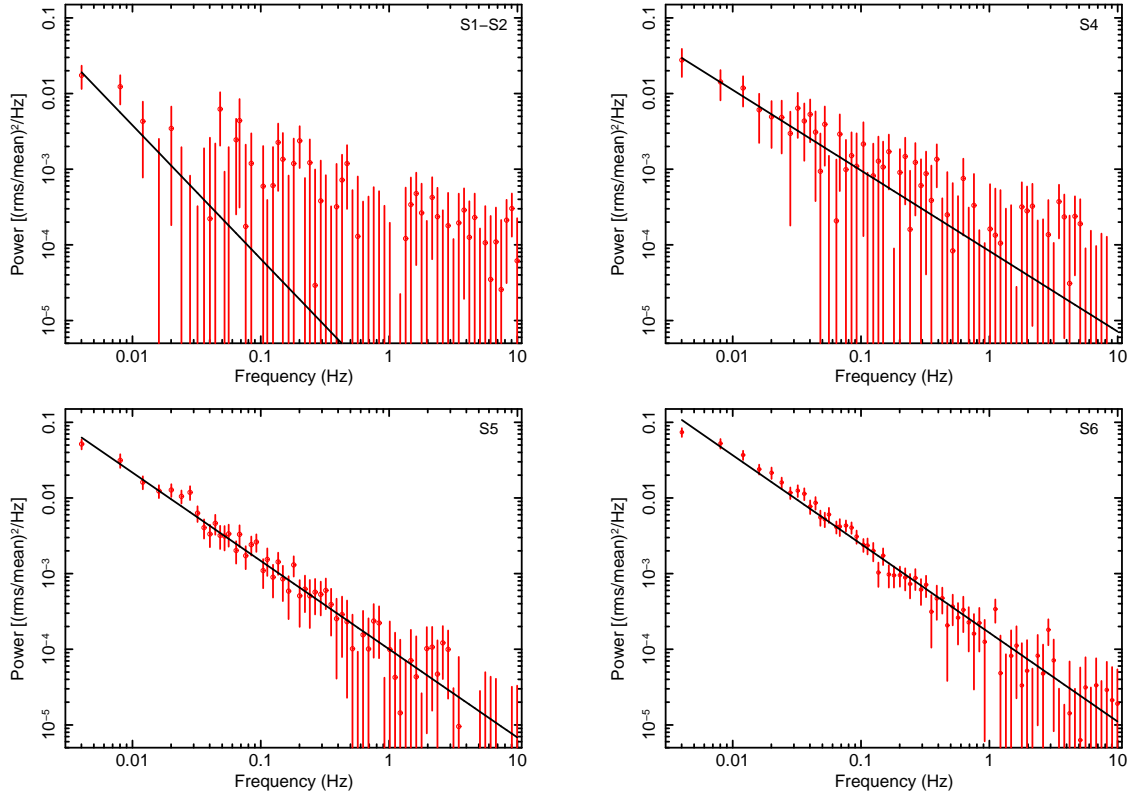

\centering
\begin{tabular}{@{}cc@{}}
\includegraphics[width=0.29\textwidth,angle=-90]{s1-s2-rms.eps}&
\includegraphics[width=0.29\textwidth, angle=-90]{s4-rms.eps} \\
\includegraphics[width=0.29\textwidth,angle=-90]{s5-rms.eps}&
\includegraphics[width=0.29\textwidth,angle=-90]{s6-rms.eps} \\
\hspace{-1in}
\end{tabular}
\caption{PSDs for different sections of the CCD. PSDs have been fitted with the power law model. See text for details.} 
\end{figure*}
\subsection{Temporal Analysis and Modeling}
\subsubsection{Timing Analysis}
We create 1.9 ms lightcurves in the energy range $3-60$ keV to investigate
the nature of Power Spectral Density (PSD). We divide the lightcurves
in the intervals of 131072 bins or size of 250 s intervals.  We rebin
the PSD in the frequency range  $0.004-260$~Hz geometrically by a factor
of 1.1 in the frequency space. We create an average PSD for a particular
section of CCD by averaging the PSD
belonging to that section. We subtract the dead time corrected Poisson
noise level \citep{Zhang95,agr18} from the PSD and  normalized
them to in units of $(rms/mean)^2/Hz$. We find that a simple power-law
(AE$^{-\alpha})$ can describe the PSD for all the sections. The signal
strength was very poor for the sections S1 and S2 and we were not be able to
constrain the parameters of the fit. Hence, we created combined PSD for
these two sections and fit the combined PSD with a power-law function.

\subsubsection{Power Spectral Properties}
PSD of all the sections of the CCD can be described by a simple power-law
function ($A\nu^{-\alpha}$). Since the signal strength is negligible
above 10~Hz, the PSD upto 10~Hz is considered for the fitting. The
parameters of the best fit are listed in Table 6. The power-law index
($\alpha$) decreases from $\sim$ 1.76 to 1.06 as the source moves from S1
to S4  and then increases as it moves from S4 to S6. The normalization
and rms (in the frequency range $0.004-50$ ~Hz) values increase as the
source moves from S1 to S6. PSDs for different sections of the CCDs are shown 
in Figure 13.

\begin{table*}
	\caption{The parameters obtained by fitting the PSD of different sections of the CCD. The rms values are in the frequency range  4~mHz to 50~Hz.}
\begin{tabular}{cccccc}
\hline
Parameters & S1-S2 & S3 & S4 & S5 & S6\\
\hline
$\alpha$          &1.76$\pm$0.31 & 1.60$\pm$0.22 & 1.06$\pm$0.05 &  1.16$\pm$0.03 & 1.17$\pm$0.02\\
$A$ ($\times$ 10$^{-5}$)& 0.11$\pm$0.62 & 0.28$\pm$0.12 & 8.35$\pm$2.62 &10.21$\pm$1.12 & 16.51$\pm$1.31\\
rms(\%) & 1.01$\pm$0.84 & 1.14$\pm$0.73 & 2.88$\pm$0.35 & 3.46$\pm$0.58 & 4.45$\pm$0.47\\
\hline
	$\chi^2/dof$ & 42/66 & 40/66 & 34/66 & 54/66 & 63/66\\
\hline
\end{tabular}
\end{table*}

\section{Discussion}
The {\it MAXI} and {\it Swift-BAT} data revealed that the source has remained in the soft spectral state since 2015 July 17 (see Figure 2). The {\it AstroSat} instruments observed this source on two occasions during the soft state. This provided a unique opportunity to study the source behaviour in detail during the soft state using broad-band and fast timing capabilities of {\it AstroSat}. The source traced a `banana' type pattern (or a C-type shape) in the CCD and hence we studied the evolution of the broad-band X-ray spectra and power spectral density (PSD) of the source along the track. We also detected two type-I X-ray bursts and investigated their properties in detail.  

The persistent spectra of the source can be well described by either a
model  consisting of  multi-colour disc emission and  cool Comptonized
component (see Figure 7) or a combination of blackbody and Comptonized emission.
In both approaches, an ionized reflection component is required at higher
flux level (CCD sections S4-S6). However, we can not distinguish between
blackbody and MCD component.  A double Comptonization model also provides
good fit to the X-ray spectra of this source. A double Comptonization
model has been previously adopted to fit the X-ray spectra of this source
\citep{thompson05, Chenvez16}.  We notice that the photon index of the
second Comptonization component could not be constrained for all the sections
while temperature was fixed at 20 keV.

We note that {\it TBabs*edge*(irefl*nthcomp+diskbb)} and {\it TBabs*edge*(irefl*nthComp+bbodyrad)}
both models provide equally good description of the data. 
Fitting the data with above two models suggests the presence of a cool ($kT_e
\sim 4.8-2.2$ keV) and an optically thick ($\tau \sim 5-21$) corona.
A cool and an optically thick Comptonized emission is generally seen
in the soft spectral states of  other Atoll sources \citep{piraino00,
barret02, Gierlinski02, tarana08, tarana07, agr18}. The study also
shows  a systematic evolution of different spectral components along the
CCD. The optical depth of the corona increases from S1 to S6. We also
note that at the same time corona also becomes cooler. Most probably, due
to increase in the supply of the seed photons along the CCD, the corona
cools down and becomes compact and more optically thick. This is
supported by the fact that supply of seed photons from disc/boundary-layer increases
(except at S2) as the source moves along the CCD. Similar variations in
the optical depth and the electron temperature have also been observed
in other Atoll sources \citep{Gierlinski02, tarana08, agr18}. We also
note that disc moves inwards as the source moves from S1 to S6. Hence,
the behaviour of the source may be related
with increase in the accretion rate along the CCD. We also note that total
flux increases from S2 to S6. However, the flux decreases from S1 to S2.

Reflection signature has been previously reported in  Z-sources GX 340+0
\citep{DAI09}, GX 17+2 \citep{agr20a} and Atoll sources 4U 1705-44
\citep{Disalvo09, Ludham17} and 4U 1636-536 \citep{Ludham17}. However,
in the case of GS 1826-238, we are reporting for the first time
the evidence of a reflection component. The source GS 1826-236 is a
high inclination ($i \sim 75^\circ$) source. In high inclination sources,
the reflection component is  difficult to observe because outer edge of disc
obscure the reflection components. The reflection component has been observed
in another high inclination ($i \sim 80-85^\circ$) NS-LMXB 4U 1822-371
\citep{Anitra21}. The reflection component observed in this source is
stronger in the high flux state, when disc extended  up to  the neutron
star surface ($10-14$ km). The investigations of other Atoll and Z-sources
also indicated that disc reflection is stronger when  accretion disc
extends close to innermost stable circular orbit (ISCO) \citep{Ludham17}.

In double Comptonization model, one Comptonized emission comes from an
optically thick corona and another comes from an optically thin corona. A
similar behaviour was observed by \cite{Chenvez16}. The total luminosity
of the source in the energy band $0.3-50.0$ keV is found to vary in the
range of  5.2 $\times$ 10$^{36} -$ 2.2$\times$ 10$^{37}$ ergs/s.

Two type-I X-ray bursts were recorded during the {\it AstroSat}
observations.  The investigation of the energy dependent lightcurves
revealed that the increase in the count rate was observed in the
hard X-rays ($20-60$ keV). Such a study has been carried out for
other Atoll sources as well and investigation shows that during the
bursts hard X-ray flux decreases.  A drop in the $30-60$ keV flux
was reported during the type-I X-ray bursts observed from Aql X-1
\citep{Maccarone03}. \cite{Chen13} also reported a shortage of $40-50$
keV flux during type-I bursts from the same source. \cite{Ji13} reported
a shortage of $30-50$ keV flux during the type I X-ray bursts observed
from the source 4U 1636-536.  A shortage of hard X-ray flux has been
reported during the type-I X-ray bursts in the source GS $1826-238$
as well \citep{Ji14,Sanchez20}.
For the first time, hard X-ray enhancement is seen during the type-I X-ray bursts detected  in the period of {\it AstroSat} observations. During the AstroSat observations, the source was in the soft state. This suggests that shortage of the hard
X-ray flux during the hard state X-ray bursts is caused by an efficient
Compton cooling of the corona. However, an additional contribution from the
burst emission to the emission of the corona along with flattening of spectra are probably causing a slight
increase in the hard X-ray flux during X-ray bursts in the soft state.
  
We investigated the spectra of burst B2 using  model {\it
TBabs(diskbb+nthComp)} and   those of B1 using model {\it TBabs(nthComp)}
(since the burst was not observed by SXT). We observed the
changes in the spectra due to the bursts. We found that photon index of Comptonized emission is
flatter during the bursts as compared to the pre-burst and post-burst
phases. However, electron temperature is lower during the X-ray bursts.
This behaviour can be explained, if during the burst Compton corona is
cooling down and increasing its optical depth. Since the corona becomes more optically
thick, number of inverse Compton scattering will increase causing
flattening of the spectra. Also, the optically thick corona will cool
down leading to decrease in electron temperature. We also note that decay
part of the spectra is softer compared to the peak part of the spectra
(see  Figure 12 and Table 5) and  disk is absent during the peak and
decay part of the bursts. Probably, very high flux from the Comptonized
corona has pushed the disk outward. The softening during the decay part
can be explained by additional cooling from the disk which is slowly
building up.  A decrease in the optical depth during the decay part can
be explained by the fact that the corona matter settles down on the disc
as it cools down. The peak luminosity of the burst B1 is 1.28$\times$
10$^{38}$ ergs/s ($2-50$ keV) and that during the burst B2 is 0.82
$\times$ 10$^{38}$ ergs/s ($2 - 50$ keV).

We also searched for the burst oscillations in the frequency range
$100-1000 $ Hz during these two bursts. No oscillations during the bursts
was observed. Finally, we also study the evolution of the PSD along the
CCD. We found that the PSD can be represented by pure power-law (Very Low Frequency Noise) and become stronger as the source moves up
along the `banana' state. A detailed timing study of Atoll source
4U 1608-522 using {\it Rossi X-ray Timing Explorer} data also revealed
that strength of very low frequency noise (VLFN) component increased as
the source moved up in the `banana' branch \citep{Vanstraa03}. They also
found narrow QPOs in this branch. The power law index for this source
varied in the range of $1.32-2.4$.  Similarly, a VLFN component which was
absent in the `island branch' appeared in the `banana' branch of Aquila X-1
\citep{Reig00}. The rms strength of VLFN component in $1-100$~Hz decreased
from 3.5 to 1.8\% as the source moved up in the `banana' branch. However,
rms strength of power-law component in the range $0.01-1$~Hz remained
almost constant (6\%).  They also found a high frequency noise component
in the `banana' branch of the source. The index of VLFN component varied
between $1.4-1.6$. A VLFN component and peaked noise were seen in the PSD
of 4U 1705-44 \citep{agr18} and  strength
of VLFN component  was not correlated with the position  on the  `banana' branch.
The index of VLFN component in GS 1826-238 varied between $1.1 - 1.8$. Therefore, we conclude that no clear trend for the rms strength and index ($\Gamma$) of the VLFN is observed in different sources as these move along the `banana' branch of their CCD. Hence, it seems that detailed behaviour of the VLFN component is source dependent.

\section*{Acknowledgements}
Authors thank the anonymous reviewer for useful and constructive suggestions which improved the quality of the paper.
VKA and AN thank GH, SAG; DD, PDMSA and Director,
URSC for encouragement and continuous support to carry out this research.
This work has used the data from the LAXPC Instruments developed at
TIFR, Mumbai and the LAXPC POC at TIFR is thanked for verifying and
releasing the data via the ISSDC data archive. We thank the AstroSat
Science Support Cell hosted by IUCAA and TIFR for providing the LaxpcSoft
software which we used for LAXPC data analysis.  This work has used the
data from the Soft X-ray Telescope ({\it SXT}) developed at TIFR, Mumbai,
and the {\it SXT} POC at TIFR is thanked for verifying \& releasing the
data and providing the necessary software tools.\\

\section*{data availability}
Data underlying this article are available at {\it AstroSat}-ISSDC website
(http://astrobrowse.issdc.gov.in/astro\_archive/archive).

\end{document}